\documentclass[pra,twocolumn,showpacs,showkeys,preprintnumbers,amsmath,amssymb]{revtex4}

\usepackage{graphicx}
\usepackage{dcolumn}
\usepackage{bm,amsmath,amsfonts,amssymb,mathrsfs,bbm}

\newcommand{\ot}[0]{\otimes}

\newcommand{\ket}[1]{|#1\rangle}
\newcommand{\bra}[1]{\langle#1|}

\newcommand{\proj}[1]{\ket{#1}\bra{#1}}

\newcommand{\be}[0]{\begin{equation}}
\newcommand{\ee}[0]{\end{equation}}
\newcommand{\bea}[0]{\begin{eqnarray}}
\newcommand{\eea}[0]{\end{eqnarray}}

\newcommand{\Tr}[0]{\mathrm{Tr}}

\begin{document}
\title{General construction of noiseless networks detecting
entanglement\\ with help of linear maps}

\author{Pawe\l{} Horodecki}
\email{pawel@mif.pg.gda.pl}
\author{Remigiusz Augusiak}
\email{remik@mif.pg.gda.pl}
\author{Maciej Demianowicz}
\email{maciej@mif.pg.gda.pl}
 \affiliation{Faculty of Applied
Physics and Mathematics, Gda\'nsk University of Technology,
Gda\'nsk, Poland}

\date{\today}

\begin{abstract}
We present the general scheme for construction of noiseless
networks detecting entanglement with the help of linear,
hermiticity-preserving maps. We show how to apply the method to
detect entanglement of unknown state without its prior
reconstruction. In particular, we prove there always exists
noiseless network detecting entanglement with the help of
positive, but not completely positive maps. Then the
generalization of the method to the case of entanglement detection
with arbitrary, not necessarily hermiticity-preserving, linear
contractions on product states is presented.
\end{abstract}

\maketitle

\section{Introduction}
It has been known that entanglement can be detected with help of
special class of maps called positive maps \cite{sep,Peres,book}.
In particular there is an important criterion \cite{sep} saying
that $\varrho$ acting on a given product Hilbert space
$\mathcal{H}_{A}\ot\mathcal{H}_{B}$ is separable if and only if
for all positive (but not completely positive) maps $\Lambda :
\mathcal{B}(\mathcal{H}_{B})\rightarrow
\mathcal{B}(\mathcal{H}_{A})$ \cite{B} the following operator
\begin{equation}
X_{\Lambda}(\varrho)=[I \otimes \Lambda](\varrho)
\end{equation}
has all non-negative eigenvalues which usually is written as
\begin{equation}
[I \otimes \Lambda](\varrho) \geq 0 \label{PositiveMaps1}.
\end{equation}
Here by $I$ we denote the identity map acting on
$\mathcal{B}(\mathcal{H}_{A})$. Since any positivity-preserving
map is also hermiticity-preserving, it makes sense to speak about
eigenvalues of $X_{\Lambda}(\varrho)$. However, it should be
emphasized that there are many $\Lambda$s (and equivalently the
corresponding criteria) and to characterize them is a hard and
still unsolved problem (see, e.g., Ref. \cite{Kossakowski} and
references therein).

For a long time the above criterion has been treated as purely
mathematical. One used to take matrix $\varrho$ (obtained in some
{\it prior} state estimation procedure) and then put it into the
formula (\ref{PositiveMaps1}). Then its spectrum was calculated
and the conclusion was drawn. However it can be seen that for, say
states acting on $\mathcal{H}_{A}\ot\mathcal{H}_{B}\sim
\mathbb{C}^{d}\ot\mathbb{C}^{d}$ and maps $\Lambda :
\mathcal{B}(\mathbb{C}^{d}) \rightarrow
\mathcal{B}(\mathbb{C}^{d})$, the spectrum of the operator
$X_{\Lambda}(\varrho)$ consists of $n_{\mathrm{spec}}=d^{2}$
elements, while full {\it prior} estimation of such states
corresponds to $n_{\mathrm{est}}=d^{4}-1$ parameters.

The question was raised \cite{PHAE} as to whether one can perform
the test (\ref{PositiveMaps1}) physically without necessity of
{\it prior} tomography of the state $\varrho$ despite the fact
that the map $I\ot\Lambda$ is not physically realizable. The
corresponding answer was \cite{PHAE} that one can use the notion
of structural physical approximation $\widetilde{I \otimes
\Lambda}$ (SPA) of un--physical map $I \otimes \Lambda$ which is
physically realizable already, but at the same time the spectrum
of the state
\begin{equation}
\tilde{X}_{\Lambda}(\varrho)=[\widetilde{I \otimes
\Lambda}](\varrho)
\end{equation}
is just an affine transformation of that of the (unphysical)
operator $X_{\Lambda}(\varrho)$. The spectrum of
$\tilde{X}_{\Lambda}(\varrho)$ can be measured with help of the
spectrum estimator \cite{Estimator}, which requires estimation of
only $d^{2}$ parameters which (because of affinity) are in one to
one correspondence with the needed spectrum of
(\ref{PositiveMaps1}). Note that for $2\ot 2$ systems (the
composite system of two qubits), similar approaches lead to the
method of detection of entanglement measures (concurrence
\cite{Concurrence} and entanglement of formation \cite{EoF})
without the state reconstruction \cite{PHPRL}.

The disadvantage of the above method is \cite{Carteret} that
realization of SPA requires addition the noise to the system (we
have to put some controlled ancillas, couple the system, and then
trace them out). In Ref. \cite{Carteret} the question was raised
about the existence of noiseless quantum networks, i.e., those of
which the only input data are: (i) unknown quantum information
represented by $\varrho^{\ot m}$ (ii) the controlled measured
qubit which reproduces us the spectrum moments (see Ref.
\cite{Estimator}). It was shown that for at least one positive map
(transposition) $T$ the noiseless network exists \cite{Carteret}.
Such networks for two-qubit concurrence and three-qubit tangle
have also been designed \cite{Carteret2}.

In the present paper we ask a general question: do noiseless
networks work only for special maps (functions) or do they exist
for any positive map test? In the case of a positive answer to the
latter: is it possible to design a general method for constructing
them? Can it be adopted to any criteria other than the one defined
in (\ref{PositiveMaps1})?

For this purpose we first show how to measure a spectrum of the
matrix $\Theta(\varrho)$, where $\Theta :
\mathcal{B}(\mathbb{C}^{m})\rightarrow\mathcal{B}(\mathbb{C}^{m})$
is an arbitrary linear, hermiticity-preserving map and $\varrho$
is a given density operator acting on $\mathbb{C}^{m}$, with the
help of only $m$ parameters estimated instead of $m^{2}-1$. For
bipartite $\varrho$ where $m=d^{2}$ this gives $d^{2}$ instead of
$d^{4}-1$. This approach is consistent with previous results
\cite{Grassl,Leifer,Brun} where arbitrary polynomials of elements
of a given state $\varrho$ have been considered. In these works it
was shown out that any at most $k$-th degree polynomial of a
density matrix $\varrho$ can be measured with help of two
collective observables on $k$ copies of $\varrho$. In fact one can
treat the moments of $\Theta(\varrho)$ which we analyze below as
polynomials belonging to such a class. We derive the explicit form
of observables for the sake of possible future application.
Moreover, approach presented in the present paper allows for quite
natural identification of observable that detects an arbitrary
polynomial of the state $\varrho$ subjected to some transformation
$\Theta$. Then we provide an immediate application in entanglement
detection showing that for suitable $\Theta$ the scheme
constitutes just a right method for detecting entanglement without
prior state reconstruction with the help of either positive map
criteria (\ref{PositiveMaps1}) or linear contraction methods
discussed later.

\section{General scheme for construction of
noiseless network detecting spectrum of $\Theta(\varrho)$}
\label{general}
\subsection{Construction of an observable}

Since $m \times m$ matrix $\Theta(\varrho)$ is hermitian its
spectrum may be calculated using only $m$ numbers
\begin{equation}\label{alfy}
\alpha_{k}\equiv \Tr
[\Theta(\varrho)]^{k}=\sum_{i=1}^{m}\lambda_{i}^{k}\qquad
(k=1,\ldots,m),
\end{equation}
where $\lambda_{i}$ are eigenvalues of $\Theta(\varrho)$. We shall
show that all these spectrum moments can be represented by mean
values of special observables. To this aim let us consider the
permutation operator $V^{(k)}$ defined by the formula
\begin{equation}\label{VKa}
V^{(k)} |e_{1}\rangle|e_{2}\rangle \otimes ... \otimes |e_{k}
\rangle= |e_{k}\rangle|e_{1}\rangle \otimes ... \otimes |e_{k-1}
\rangle,
\end{equation}
where $(k=1,\ldots,m)$ and $\ket{e_{i}}$ are vectors from
$\mathbb{C}^{m}$. One can see that $V^{(1)}$ is just an identity
operator $\mathbbm{1}_{m}$ acting on $\mathbb{C}^{m}$. Combining
Eqs. (\ref{alfy}) and (\ref{VKa}) we infer that $\alpha_{k}$ may
be expressed by relation
\begin{equation}
\alpha_{k}=\Tr \left\{V^{(k)}[\Theta(\varrho)]^{\otimes k}\right\}
\label{alfa}
\end{equation}
which is generalization of the formula from Refs.
\cite{Estimator,PHAE} where $\Theta$ was (unlike here) required to
be a physical operation. At this stage the careful analysis  of
the right--hand side of Eq. (\ref{alfa}) shows that $\alpha_{k}$
is a polynomial of at most $k$-th degree in matrix elements of
$\varrho$. This, together with the observation of Refs.
\cite{Brun,Grassl,Leifer} allows us already to construct a single
collective observable that detects $\alpha_{k}$. However, for the
sake of possible future applications we derive the observable
explicitly below. To this aim we first notice that $\alpha_{k}$
may be obtained using hermitian conjugation of $V^{(k)}$ which
again is a permutation operator but permutes states $\ket{e_{i}}$
in the reversed order. Therefore all the numbers $\alpha_{k}$ may
be expressed
as
\begin{equation}\label{alfa2}
\alpha_{k}=\frac{1}{2}\Tr\left[\left(V^{(k)}+V^{(k)\dagger}\right)\Theta(\varrho)^{\otimes
k}\right].
\end{equation}
Let us focus for a while on the map $\Theta$. Due to its
hermiticity-preserving property it may be expressed as
\begin{equation}
\Theta(\cdot)=\sum_{j=0}^{m^{2}-1}\eta_{j}K_{j}(\cdot)K_{j}^{\dagger}
\end{equation}
with $\eta_{j}\in\mathbb{R}$ and $K_{j}$ being linearly
independent $m$-by-$m$ matrices. By the virtue of this fact and
some well-known properties of the trace, after rather
straightforward algebra we may rewrite Eq. (\ref{alfa2}) as
\begin{equation}\label{alfa3}
\alpha_{k}=\frac{1}{2}\Tr\left[\left(\Theta^{\dagger}\right)^{\ot
k}\left(V^{(k)}+V^{(k)\dagger}\right)\varrho^{\ot k}\right],
\end{equation}
where $\Theta^{\dagger}$ is a dual map to $\Theta$ and is given by
$\Theta^{\dagger}(\cdot)=\sum_{i}\eta_{i}K_{i}^{\dagger}(\cdot)K_{i}$.
Here we have applied a map $(\Theta^{\dagger})^{\ot k}$ on the
operator $V^{(k)}+V^{(k)\dagger}$ instead of applying $\Theta^{\ot
k}$ to $\varrho^{\otimes k}$. This apparently purely mathematical
trick with the aid of the fact that the square brackets in the
above contain a hermitian operator allows us to express the
numbers $\alpha_{k}$ as a mean value of some observables in the
state $\varrho^{\ot k}$. Indeed, introducing
\begin{equation}\label{obs}
\mathcal{O}^{(k)}_{\Theta}=
\frac{1}{2}\Tr\left[\left(\Theta^{\dagger}\right)^{\ot
k}\left(V^{(k)}+V^{(k)\dagger}\right)\right]
\end{equation}
we arrive at
\begin{equation}
\alpha_{k}=\Tr\left[\mathcal{O}^{(k)}_{\Theta} \varrho^{\otimes
k}\right]. \label{MeanValues}
\end{equation}

In general, a naive measurement of all mean values would require
estimation of much more parameters that $m$. But there is a
possibility of building a unitary network that requires estimation
of exactly $m$ parameters using the idea that we recall and refine
below.

Finally, let us notice that the above approach generalizes
measurements of polynomials of elements of $\varrho$ in the sense
that it shows explicitly how to measure the polynomials of
elements of $\Theta(\varrho)$. Of course, this is only of rather
conceptual importance since both issues are mathematically
equivalent and have the origin in Refs. \cite{Grassl,Leifer,Brun}.

\subsection{Detecting mean of an observable by measurement on a
single qubit revised}
Let $\mathcal{A}$ be an arbitrary observable (it may be even
infinite dimensional) which spectrum lies between finite numbers
$a^{\min}_{\mathcal{A}}$ and $a^{\max}_{\mathcal{A}}$ and $\sigma$
be a state acting on $\mathcal{H}$. In Ref. \cite{binpovm} it has
been pointed out that the mean value $\langle \mathcal{A}
\rangle_{\sigma}= \Tr\mathcal{A}\sigma$ may be estimated in
process involving the measurement of only one qubit. This fact is
in good agreement with further proof that single qubits may serve
as interfaces connecting quantum devices \cite{Lloyd}. Below we
recall the mathematical details of the measurement proposed in
Ref. \cite{binpovm}. At the beginning one defines the following
numbers
\begin{equation}
a^{(-)}_{\mathcal{A}}\equiv
\max\{0,-a^{\min}_{\mathcal{A}}\},\qquad a^{(+)}_{\mathcal{A}}
\equiv a^{(-)}_{\mathcal{A}}+a^{\max}_{\mathcal{A}},
\end{equation}
and observe that the hermitian operators
\begin{eqnarray}
V_{0}=\sqrt{\left(a^{(-)}_{\mathcal{A}}\mathbbm{1}_{\mathcal{H}}+\mathcal{A}\right)\Big/a^{(+)}_{\mathcal{A}}}
\end{eqnarray}
and
\begin{equation}
V_{1}=\sqrt{\mathbbm{1}_{\mathcal{H}} - V_{0}^\dagger V_{0}}
\end{equation}
satisfy
$\sum_{i=0}^{1}V_{i}^{\dagger}V_{i}=\mathbbm{1}_{\mathcal{H}}$
\cite{Identity} and as such define a generalized quantum
measurement which can easily be extended to a unitary evolution
(see Appendix A of Ref. \cite{APS} for a detailed description).
Consider a partial isometry on the Hilbert space $\mathbb{C}^{2}
\otimes \mathcal{H}$ defined by the formula
\begin{equation}
\tilde{U}_{\mathcal{A}}=\sum_{i=0}^{1} |i\rangle \langle 0|
\otimes V_{i}=\left(
\begin{array}{cc}
V_{0} & 0\\
V_{1} & 0
\end{array} \right).
\end{equation}
The first Hilbert space $\mathbb{C}^{2}$ represents the qubit
which shall be measured in order to estimate the mean value
$\langle \mathcal{A} \rangle_{\sigma}$. The partial isometry can
always be extended to unitary $U_{\mathcal{A}}$ such that if it
acts on $|0 \rangle \langle 0| \otimes \sigma$ then the final
measurement of observable $\sigma_{z}$ \cite{Pauli} on the first
(qubit) system gives probabilities "spin-up" (of finding it in the
state $|0\rangle$) and "spin-down" (of finding in state
$|1\rangle$), respectively of the form
\begin{equation}
p_{0}=\Tr\left(V_{0}^\dagger V_{0}\varrho\right), \qquad
p_{1}=\Tr\left(V_{1}^\dagger V_{1}\varrho\right)=1-p_{0}.
\end{equation}
One of the possible extensions of $\tilde{U}_{\mathcal{A}}$ to the
unitary on $\mathbb{C}^{2}\ot \mathcal{H}$ is the following
\begin{equation}\label{isometry}
U_{\mathcal{A}}=\left(
\begin{array}{cc}
V_{0} & -V_{1}\\
V_{1} & V_{0}
\end{array}
\right)=\mathbbm{1}_{2}\ot V_{0}-i\sigma_{y}\ot V_{1}.
\end{equation}
The unitarity of $U_{\mathcal{A}}$ follows from the fact that
operators $V_{0}$ and $V_{1}$ commute. Due to the practical
reasons instead of unitary operation representing POVM
$\{V_{0},V_{1}\}$ we shall consider
\begin{equation}\label{Udet}
U^{\mathrm{det}}(\mathcal{A},U'_{\mathcal{H}})=\left(\mathbbm{1}_{2}
\ot U_{\mathcal{H}}'\right)U_{\mathcal{A}}\left(\mathbbm{1}_{2}\ot
U_{\mathcal{H}}'\right)^{\dagger},
\end{equation}
where $\mathbbm{1}_{2}$ is an identity operator on the one-qubit
Hilbert space $\mathbb{C}^{2}$ and $U_{\mathcal{H}}'$ is an
arbitrary unitary operation that acts on ${\cal H}$ and simplifies
the decomposition of $U_{\mathcal{A}}$ into elementary gates. Now
if we define a mean value of measurement of $\sigma_{z}$ on the
first qubit after action of the network (which sometimes may be
called visibility):
\begin{equation}
v_{\mathcal{A}}=\Tr\left[ \left(\sigma_{z}\otimes
\mathbbm{1}_{\mathcal{H}}\right) \left(\mathbbm{1}_{2} \otimes
U_{\mathcal{H}}'\right) U_{\mathcal{A}} \mathcal{P}_{0}\ot\sigma
U^{\dagger}_{\mathcal{A}}\left(\mathbbm{1}_{2} \otimes
U_{\mathcal{H}}'\right)^{\dagger}\right], \label{vis}
\end{equation}
where $\mathcal{P}_{0}$ is a projector onto state $\ket{0}$, i.e.,
$\mathcal{P}_{0}=\proj{0}$, then we have an easy formula for the
mean value of the initial observable $\mathcal{A}$:
\begin{equation}\label{meanA}
\langle \mathcal{A}\rangle_{\sigma}=a^{(+)}_{\mathcal{A}}p_{0}-
a^{(-)}_{\mathcal{A}}=a^{(+)}_{\mathcal{A}}\frac{v_{\mathcal{A}}+1}{2}-a^{(-)}_{\mathcal{A}}.
\end{equation}

\begin{figure}
\centerline{\includegraphics[width=8cm]{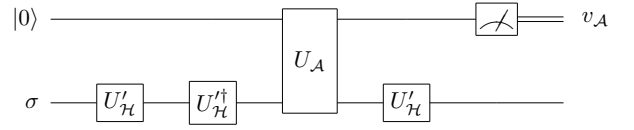}}
\caption{General scheme of a network for estimating mean value of
an observable $\mathcal{A}$, with a bounded spectrum, in a given
state $\sigma$. Both $U_{\mathcal{H}'}$ and its conjugate
$U_{\mathcal{H}}'^{\dagger}$ standing before $U_{\mathcal{A}}$ can
obviously be removed as they give rise to identity, last unitary
on the bottom wire can be removed as it does not impact
measurement statistics on the top qubit. However, they have been
put to simplify subsequent network structure.} \label{Fig1}
\end{figure}

A general scheme of a network estimating the mean value
(\ref{meanA}) is provided in Fig. \ref{Fig1}.
%
%
We put an additional unitary operation on the bottom wire after
unitary $U_{\mathcal{A}}$ (which does not change the statistics of
the measurement on control qubit) and divided identity operator
into two unitaries acting on that wire which explicitly shows how
simplification introduced in Eq. (\ref{Udet}) works in practice.

Now one may ask if the mean value
$\langle\mathcal{A}\rangle_{\sigma}$ belongs to some fixed
interval, i.e.,
\begin{equation}\label{int}
c_{1}\le\langle\mathcal{A}\rangle_{\sigma}\le c_{2},
\end{equation}
where $c_{1}$ and $c_{2}$ are real numbers belonging to the
spectrum of $\mathcal{A}$, i.e.,
$[a_{\mathcal{A}}^{\mathrm{min}},a_{\mathcal{A}}^{\mathrm{max}}]$
(e.g. if $\mathcal{A}$ is an entanglement witness and we want to
check the entanglement of a state $\sigma$ then we can put
$c_{1}=0$ and $c_{2}=a_{\mathcal{A}}^{\mathrm{max}}$, and
condition (\ref{int}) reduces to
$\langle\mathcal{A}\rangle_{\sigma}\ge 0$). Then one easily infers
that the condition (\ref{int}) rewritten for visibility is
\begin{equation}
2\frac{c_{1}+a_{\mathcal{A}}^{(-)}}{a_{\mathcal{A}}^{(+)}}-1\le
v_{\mathcal{A}}\le
2\frac{c_{2}+a_{\mathcal{A}}^{(-)}}{a_{\mathcal{A}}^{(+)}}-1.
\end{equation}

\begin{figure}
\includegraphics[width=8.5cm]{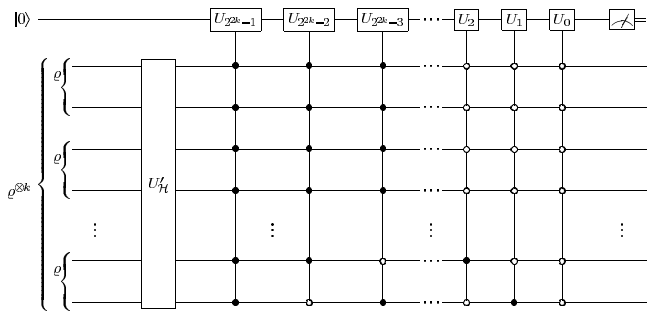}
\caption{Noiseless network for estimating moments of
$\Theta(\varrho)$ with $\varrho$ being a bipartite mixed state,
i.e., density matrix acting on $\mathbb{C}^{2}\ot\mathbb{C}^{2}$.}
\label{Fig2}
\end{figure}

Having the general network estimating $v_{\mathcal{A}}$, one needs
to decompose an isometry $U_{\mathcal{A}}$ onto elementary gates.
One of possible ways to achieve this goal is, as we shall see
below, to diagonalize the operator $V_{0}$. Hence we may choose
$U_{\mathcal{H}}'$ (see Eq. (\ref{Udet})) to be
\begin{equation}
U_{\mathcal{H}}'=\sum_{\bold{k}}
\ket{{\bold{k}}}\bra{\phi_{\bold{k}}}
\end{equation}
with $\ket{\phi _{\bold{k}}}$ being normalized eigenvectors of
$V_{0}$ indexed by a binary number with length $2^k$. Since
$V_{0}$ and $V_{1}$ commutes, this operation diagonalizes $V_{1}$
as well. By virtue of these facts, Eq. (\ref{Udet}) reduces to
\begin{equation}
U^{\mathrm{det}}(\mathcal{A},U_{\mathcal{H}'})=\sum_{\bold{k}}
U_{\bold{k}}\otimes\proj{\bold{k}},
\end{equation}
with unitaries (as previously indexed by a binary number)
\begin{equation}
U_{\bold{k}}=\sqrt{\lambda_{\bold{k}}}\mathbbm{1}_{2}-i\sqrt{1-\lambda_{\bold{k}}}\sigma_y,
\end{equation}
where $\lambda_{\bold{k}}$ are eigenvalues of $V_{0}$. So in fact
we have a combination of operations on the first qubit controlled
by $2^k$ wires. All this combined gives us the network shown in
the Fig. \ref{Fig2}.

Now we are in the position to combine all the elements presented
so far an show how, if put together, they provide the general
scheme for constructing noiseless network for spectrum of
$\Theta(\varrho)$ for a given quantum state $\varrho$. For the
sake of clarity below we itemize all steps necessary to obtain the
spectrum of $\Theta(\varrho)$:
\begin{description}
    \item[(i)] Take all observables $\mathcal{O}^{(k)}$ $(k=1,\ldots,m)$ defined by Eq. (\ref{obs}).
    \item[(ii)] Construct unitary operations
$U_{\mathcal{O}^{(k)}}$ according to the the given prescription.
Consider the unitary operation
$U^{\mathrm{det}}(\mathcal{A},U'_{\mathcal{H}})$
($U'_{\mathcal{H}}$ arbitrary). Find decomposition of the
operation into elementary quantum gates and minimize the number of
gates in the decomposition with respect to $U_{\mathcal{H}}'$.
Build the (optimal) network found in this way.
    \item[(iii)] Act with the network on initial state
    $\mathcal{P}_{0} \otimes \varrho^{\otimes k}$.
    \item[(iv)] Measure the ,,visibilities''
$v_{\mathcal{O}^{(k)}_{\Theta}}$ $(k=1,\ldots,m)$ according to
(\ref{vis}).
    \item[(v)] Using Eq. (\ref{meanA}) calculate the values of
    $\alpha_{k}$ $(k=1,\ldots,m)$
representing the moments of $\Theta(\varrho)$.
\end{description}
%

\subsection{Detecting entanglement with networks: example}
The first obvious application of the presented scheme is
entanglement detection {\it via} positive but not completely
positive maps. In fact for any bipartite  state
$\varrho\in\mathcal{B}(\mathcal{H}_{A}\ot\mathcal{H}_{B})$ we only
need to substitute $\Theta$ with $\mathbbm{1}_{A} \otimes
\Lambda_{B}$ with $\Lambda_{B}$ being some positive map. Then
application of the above scheme immediately reproduces all the
results of the schemes from Ref. \cite{PHAE} but without
additional noise added (presence of which required more precision
in measurement of visibility).

\begin{figure}[h]
\centerline{\includegraphics[width=8.5cm]{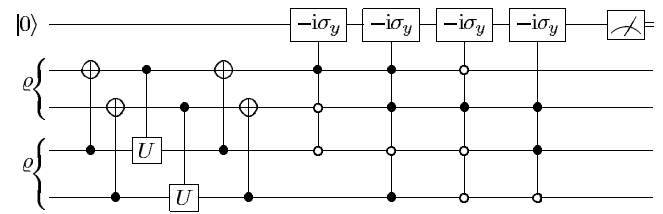}}
\caption{Network estimating the second moment of partially
transposed two-qubit density matrix $\varrho$. $U_{\mathcal{H}'}$
is decomposed to single qubit gates; here $\displaystyle
U=(1/\sqrt{2})(\mathbbm{1}_{2}+i\sigma_{y})$. } \label{Fig3}
\end{figure}

As an illustrative example consider $\Lambda_{B}=T$, i.e.,
$\Theta$ is partial transposition on the second subsystem (usually
denoted by $T_B$ or by $\Gamma$), in $2\otimes 2$ systems. Due to
to the fact that partial transposition is trace--preserving we
need only three numbers $\alpha _{k}$, ($k=2,3,4$) measurable {\it
via} observables
\begin{equation}
\mathcal{O}^{(2)}_{T}=V_1^{(2)}\otimes V_2^{(2)}
\end{equation}
and
\begin{equation}
\mathcal{O}^{(3,4)}_{T}=\frac{1}{2}\left( V_1^{(3,4)}\otimes
V_2^{(3,4)\dagger}+V_1^{(3,4)\dagger}\otimes V_2^{(3,4)}\right),
\end{equation}
where subscripts mean that we exchange first and second subsystems
respectively. The hermitian conjugation in the above may be
replaced by transposition since the permutation operators have
real entries. For simplicity we show only the network measuring
second moment of $\varrho^{T_{B}}$. General scheme from Fig.
\ref{Fig2}. reduces then to the scheme from Fig. \ref{Fig3}.
Note that the network can also be regarded as a one measuring
purity of a state as $\Tr(\varrho ^{T_{B}})^2=\Tr\varrho ^2$. Note
that the this network is not optimal since an alternative network
\cite{Estimator} measuring $\Tr\varrho^{2}$ requires two
controlled swaps.

\section{Extension to linear contractions criteria}
The above approach may be generalized to the so-called {\it linear
contractions criteria}. To see this let us recall that the
powerful criterion called computable cross norm (CCN) or matrix
realignment criterion has recently been introduced
\cite{CCN,CCN1}. This criterion is easy to apply (involves simple
permutation of matrix elements) and has been shown \cite{CCN,CCN1}
to be independent on a positive partial transposition (PPT) test
\cite{Peres}. It has been further generalized to the {\it linear
contractions criterion} \cite{PartialCCN} which we shall recall
below. If by $\varrho_{A_{i}}\;(i=1,\ldots,n)$ we denote density
matrices acting on Hilbert spaces $\mathcal{H}_{A_{i}}$ and by
$\tilde{\mathcal{H}}$ certain Hilbert space, then for some linear
map $\mathcal{R} :
\mathcal{B}(\mathcal{H}_{A_{1}}\ot\ldots\ot\mathcal{H}_{A_{n}})\rightarrow
\mathcal{B}(\tilde{\mathcal{H}})$ we have the following

{\bf Theorem} \cite{PartialCCN}. {\it If some ${\cal R}$ satisfies
\begin{equation}\label{Theorem}
\left|\left|{\mathcal {R}}\left(\varrho_{A_1} \ot \varrho_{A_2}
\ot \ldots \ot \varrho_{A_n}\right)\right|\right|_{\Tr}\leq 1,
\end{equation}
then for any separable state $\varrho_{A_1A_2 \ldots
A_n}\in\mathcal{B}(\mathcal{H}_{A_{1}}\ot\ldots\ot\mathcal{H}_{A_{n}})$
one has}
\begin{equation}\label{Theorem2}
||\mathcal {R}(\varrho_{A_1A_2 \ldots A_n})||_{\Tr}\leq 1.
\end{equation}
The maps $\mathcal{R}$ satisfying (\ref{Theorem}) are linear
contractions on product states and hereafter they shall be called,
in brief, linear contractions. In particular, the separability
condition (\ref{Theorem2}) comprises the generalization of the
realignment test to permutation criteria \cite{PartialCCN,Chen}
(see also Ref. \cite{Fan}).

The noisy network for entanglement detection with the help of the
latter have been proposed in Ref. \cite{PHPLA2003}. Here we
improve this result in two ways, namely, by taking into account
all maps $\mathcal{R}$ of type (\ref{Theorem}) (not only
permutation maps) and introducing the corresponding noiseless
networks instead of noisy ones. For these purposes we need to
generalize the lemma from Ref. \cite{PHPLA2003} formulated
previously only for real maps $\mathcal{S} :
\mathcal{B}(\mathcal{H})\rightarrow\mathcal{B}(\mathcal{H})$. We
represent action of $\mathcal{S}$ on any $\varrho\in
\mathcal{B}(\mathcal{H})$ as
\begin{equation}
{\mathcal{S}}(\varrho)=\sum_{ij,kl}{\mathcal{S}}_{ij,kl}\Tr(\varrho
P_{ij}) P_{kl},
\end{equation}
where in Dirac notation $P_{xy}=|x\rangle \langle y|$. Let us
define complex conjugate of the map $\mathcal{S}$ {\it via}
complex conjugation of its elements, i.e.,
\begin{equation}
{\mathcal{S}}^{*}(\varrho)=\sum_{ij,kl}{\mathcal{S}}_{ij,kl}^{*}\Tr(\varrho
P_{ij}) P_{kl},
\end{equation}
where asterisk stands for the complex conjugation. The we have the
following lemma which is easy to proof by inspection:

{\bf Lemma.} {\it Let ${\mathcal{S}}$ be an arbitrary linear map
on $\mathscr{B}(\mathcal{H})$. Then the map ${\mathcal{S}}' \equiv
[T \circ {\mathcal{S}}^{*} \circ T]$ satisfies ${\mathcal{S}}'
(\varrho)=[{\mathcal{S}}(\varrho) ]^{\dagger}$.}

Now let us come to the initial problem of this section. Suppose
then we have $\mathcal{R}$ satisfying Eq. (\ref{Theorem}) and a
given physical source producing copies of a system in state
$\varrho$ for which we would like to check Eq. (\ref{Theorem2}).
Let us observe that
\begin{equation}\label{24}
||\mathcal{R}(\varrho)||_{\mathrm{Tr}}=\sum_{i}\sqrt{\gamma_{i}},
\end{equation}
where $\{\gamma_{i}\}$ are eigenvalues of the operator
%
$X_{\mathcal{R}}(\varrho)=\mathcal{R}(\varrho)\mathcal{R}(\varrho)^{\dagger}.$
%
Below we show how to find the spectrum $\{\gamma_{i}\}$. We need
to apply our previous scheme from Sec. \ref{general} to the
special case. Let us define the map $L_{\mathcal{R}}=\mathcal{R}
\otimes {\cal R}'$, where ${\cal R}$ is our linear contraction and
$\mathcal{R}'$ is defined according to the prescription given in
the Lemma above, i.e., $\mathcal{R}'=[T\circ \mathcal{R}^{*}\circ
T]$. Let us also put $\varrho'=\varrho^{\ot 2}$ and apply the
scheme presented above to detect the spectrum of $L_{\cal
R}(\varrho')$. It is easy to see that the moments detected in that
way are
\begin{equation}
\Tr [L_{\mathcal{R}}(\varrho')]^{k}=
\Tr\left[\mathcal{R}(\varrho)\mathcal{R}(\varrho)^{\dagger}\right]^{k}=\sum_{i}\gamma_{i}^{k}.
\end{equation}
From the moments one easily reconstructs $\{\gamma_{i}\}$ and may
check the violation of Eq. (\ref{Theorem2}).

\section{Summary}
\label{Summary}

We have shown how to detect the spectrum of the operator
$\Theta(\varrho)$ for arbitrary linear hermiticity-preserving map
$\Theta$ given the source producing copies of the system in state
$\varrho$. The network involved in the measurement is noiseless in
the sense of \cite{Carteret} and the measurement is required only
on the controlled qubit. Further we have shown how to apply the
method to provide general noiseless network scheme of detection
detecting entanglement with the help of criteria belonging to one
of two classes, namely, those involving positive maps and applying
linear contractions on product states.

The structure of the proposed networks is not optimal and needs
further investigations. Here however we have been interested in
quite a fundamental question which is interesting by itself: {\it
Is it possible to get noiseless networks schemes for any criterion
from one of the above classes?} Up to now their existence was
known {\it only} for special case of positive partial transpose
(cf. \cite{Carteret2}). Here we have provided a positive answer to
the question.

Finally, let us note that the above approach can be viewed as an
application of collective observables [see Eq.
(\ref{MeanValues})]. The general paradigm initiated in Refs.
\cite{PHPRL,PHPRA2003} has been recently fruitfully applied in the
context of general concurrence estimates
\cite{AolitaMintert,MintertBuchleitner} which has been even
preliminarily experimentally illustrated. Moreover, recently the
universal collective observable detecting any two-qubit
entanglement has been constructed \cite{My}. It seems that the
present approach needs further analysis from the point of view of
collective observables including especially collective
entanglement witness (see \cite{PHPRA2003,MintertBuchleitner}).

 \acknowledgments P. H. thanks Artur Ekert for
valuable discussions. The work is supported by the Polish Ministry
of Science and Education under the grant No. 1 P03B 095 29, EU
project QPRODIS (IST-2001-38877) and IP project SCALA. Figures
were prepared with help of QCircuit
package.


\begin{thebibliography}{0}
\bibitem{Identity}By
$\mathbbm{1}_{\mathcal{H}}$ we denote an identity operator on
$\mathcal{H}$.
\end{thebibliography}
\end{document}